\documentclass[preprint, superscriptaddress, showpacs,preprintnumbers,amsmath,amssymb,prd]{revtex4}
\usepackage{graphicx}

\begin{document}

\thispagestyle{empty}

\title{Constraints on axion-nucleon coupling constants
from measuring the Casimir force between corrugated surfaces
}

\author{V.~B.~Bezerra}
\affiliation{Department of Physics, Federal University of Para\'{\i}ba,
C.P.5008, CEP 58059--970, Jo\~{a}o Pessoa, Pb-Brazil}
\author{
G.~L.~Klimchitskaya}
\affiliation{Central Astronomical Observatory
at Pulkovo of the Russian Academy of Sciences,
St.Petersburg, 196140, Russia}
\affiliation{Institute of Physics, Nanotechnology and
Telecommunications, St.Petersburg State
Polytechnical University, St.Petersburg, 195251, Russia}
\affiliation{Department of Physics, Federal University of Para\'{\i}ba,
C.P.5008, CEP 58059--970, Jo\~{a}o Pessoa, Pb-Brazil}
\author{
 V.~M.~Mostepanenko}
\affiliation{Central Astronomical Observatory
at Pulkovo of the Russian Academy of Sciences,
St.Petersburg, 196140, Russia}
\affiliation{Institute of Physics, Nanotechnology and
Telecommunications, St.Petersburg State
Polytechnical University, St.Petersburg, 195251, Russia}
\affiliation{Department of Physics, Federal University of Para\'{\i}ba,
C.P.5008, CEP 58059--970, Jo\~{a}o Pessoa, Pb-Brazil}
\author{C.~Romero}
\affiliation{Department of Physics, Federal University of Para\'{\i}ba,
C.P.5008, CEP 58059--970, Jo\~{a}o Pessoa, Pb-Brazil}

\begin{abstract}
We obtain stronger laboratory constraints on
the coupling constants of axion-like particles to
nucleons from measurements of the normal and lateral
Casimir forces between sinusoidally corrugated surfaces
of a sphere and a plate. For this purpose, the normal
and lateral additional force arising in the experimental
configurations due to two-axion exchange between protons
and neutrons are calculated. Our constraints following
from measurements of the normal and lateral
Casimir forces are stronger than the laboratory
constraints reported so far for masses of
axion-like particles larger
than 11\,eV and 8\,eV, respectively.
A comparison between various
laboratory constraints on the
coupling constants of axion-like particles to nucleons
obtained from the magnetometer
measurements, E\"{o}tvos- and Cavendish-type experiments,
and from the Casimir effect is performed over the
wide range of masses of axion-like particles
from $10^{-10}\,$eV to 20\,eV.
\end{abstract}

\maketitle
\section{Introduction}

After the publication of the papers by Peccei and Quinn \cite{1},
Weinberg \cite{2} and Wilczek \cite{3}, light pseudoscalar
particles named axions have become the subject of active
theoretical and experimental investigations.
This is because there is no other natural explanation for
the absence of large {\it CP} violation and respective
electric dipole moment of the neutron in strong interactions
predicted by quantum chromodynamics (QCD).
In addition, axions reasonably explain the nature of dark
matter in astrophysics \cite{4,5} which makes them of prime
importance for our general physical concepts.

A lot of experiments on searching axions have been performed
(see Refs.~\cite{6,7,8,9,10} for a review), but the
detection was not succeeded yet.
After the originally introduced QCD axions were
constrained to a very narrow band of parameter space, many
kinds of axion-like particles have been discussed in the
framework of Grand Unified Theory (GUT) and string theory
\cite{10}. Specifically, much attention has been given to the models
of GUT (DFSZ) axions \cite{11,12} and the so-called {\it hadronic}
(KSVZ) axions \cite{13,14}. The latter model
involves a relationship
between the axion-nucleon coupling constant and the breaking
scale of the Peccei-Quinn symmetry and, in its turn, falls into
several groups of models (see, e.g., Refs.~\cite{15,16}).

Axion-like particles can interact with photons, electrons and nucleons.
These interactions are used for experimental searches of
axion-like particles in both the laboratory experiments
(see below) and in astrophysics (see,
for instance, constraints on the axion-photon and axion-electron
coupling constants obtained by means of axion solar telescope
\cite{17} and from the Compton process and electron-positron
annihilation with an axion emission in stellar plasmas
\cite{18,19}, respectively).
Both the laboratory and astrophysical
constraints may depend on the model
of axions used
containing different relations between couplings to
different particles
(for instance, the constraints from the solar
axion data \cite{20,21} and from the neutrino data of
supernova SN~1987A \cite{22} are found using the model of
hadronic axions, where the coupling constant is a function of
the axion mass).
The astrophysical constraints may also depend on complicated
behavior of matter in stars.
For example, the coupling constants
of hadronic axions to nucleons, obtained \cite{23,24}
from an axion emission
rate during stellar cooling, depend \cite{24} on
significant uncertainties in the dense nuclear matter effects.
Furthermore, the astrophysical constraints could depend on
the environment \cite{24a,24b,24c}.
This is the reason why the
constraints following from the table-top laboratory experiments,
which avoid such kind uncertainties
and applicable to a wide range of axion-like particles, are of much
interest for axion physics.
At the same time it is necessary to stress that the astrophysical
constraints are often much stronger than the laboratory limits.
Thus, from the solar axion data the upper limits for axion mass
$m_a\leq 159\,$eV \cite{20} and $m_a\leq 145\,$eV \cite{21}
were obtained. {}From the neutrino data of supernova SN 1987A and
from stellar cooling by the emission of hadronic axions the
coupling constant of axions to nucleons was found to be less or
of order $10^{-10}$ \cite{22,23,24}.

In Ref.~\cite{25} the model-independent constraints on an
axion-neutron coupling constants were found with the help of a
magnetometer using spin-polarized atoms.
These constraints are the strongest for $m_a$ from $10^{-10}$
to $6\times 10^{-6}\,$eV.
In Ref.~\cite{26} the constraints on an
axion-nucleon coupling constants were obtained from the laboratory
 experiments of E\"{o}tvos \cite{27,28} and Cavendish \cite{29,30}
 type for $m_a$ from $10^{-9}$ to $10^{-5}\,$eV.
Keeping in mind that the test bodies in these experiments were
unpolarized, any additional force could come from the two-axion
exchange between nucleons belonging to different test bodies.
Using the same approach, stronger constraints on the
axion-nucleon coupling constants were obtained in Ref.~\cite{31}
from the results of more modern Cavendish-type experiment
\cite{32}
in the region of axion masses from $10^{-6}$ to $10^{-2}\,$eV.
The additional force arising due to two-axion exchange between
nucleons was then used \cite{33,34,35} to constrain the
axion-nucleon coupling constants from experiments on measuring
the Casimir-Polder \cite{36} and Casimir
\cite{37,38,39,40,41,42,43} interactions.
This allowed to strengthen the laboratory constraints on
the coupling constants of axion-like particles to nucleons in the
wide range of their masses from $10^{-3}\,$eV to 15\,eV.

The previously obtained constraints on axion-like particles from
the Casimir effect \cite{33,34,35} used experiments with either
plane or spherical boundary surfaces. There is, however, another
class of experiments on measuring the Casimir force exploiting
test bodies with corrugated surfaces. The normal Casimir force
between a smooth Au-coated sphere and a rectangular corrugated
Si plate was measured and found in agreement with the scattering
theory in Refs.~\cite{44,45,46}.  The use of Si, whose density
is smaller than of Au, does not allow to get strongest
constraints on axions using the data of these experiments.
The normal Casimir force between sinusoidally corrugated
Au-coated surfaces of a sphere and a plate was measured in
Ref.~\cite{47} and also found in agreement with computations
using the scattering theory. This experiment is rather
promising for constraining the coupling constants of axion-like
particles to nucleons.
In the experiments of Refs.~\cite{48,49} the lateral Casimir
force acting in tangential direction to the sinusoidally
corrugated Au-coated surfaces of a sphere and a plate has been
measured. The comparison with computational results using the
scattering theory demonstrated a very good agreement with the
measurement data \cite{48,49}. Thus, these experiments are also
of high promise for obtaining new constraints on the  interaction
of axion-like particles with nucleons.

In this paper, we obtain the constraints on
the coupling constants of axion-like particles to nucleons
using the measure of agreement with theory of the experimental
data of Refs.~\cite{47,48,49} exploiting the corrugated
Au-coated surfaces of a sphere and a plate.
For this purpose, we calculate the additional normal and
lateral forces arising between the sinusoidally corrugated
surfaces of a sphere and a plate due to two-axion exchange.
The constraints are found from the fact that these forces were
not observed in the limits of experimental error.
They are shown to be the strongest laboratory
constraints in the region of masses of
axion-like particles from 8 to 20\,eV.
This conclusion is obtained from the comparison with other
laboratory constraints on the coupling constants of
axion-like particles to nucleons obtained in the literature
for $m_a$ larger than $10^{-10}\,$eV.

The paper is organized as follows. In Sec.~II we calculate the
additional normal force due to two-axion exchange between the
sinusoidally corrugated bodies and derive the constraints on
the coupling constants of  axion-like particles
to nucleons from the experimental data
of Ref.~\cite{47}. Section~III contains the derivation of the
additional lateral force due to  exchange of two axion-like
particles.
The respective constraints are found from the measurement
data of Refs.~\cite{48,49}. In Sec.~IV we present the
comparison of our results with other laboratory
constraints obtained in the literature. Section~V contains our
conclusions and discussion.

Throughout the paper we use units in which $\hbar=c=1$.

\section{Constraints from measuring the normal Casimir force
between sinusoidally corrugated surfaces}

In Ref.~\cite{47} the normal Casimir force was measured between
a sinusoidally corrugated Au-coated plate and a sinusoidally
corrugated Au-coated sphere at various angles between the uniaxial
corrugations from $0^{\circ}$ to $2.4^{\circ}$ using an atomic
force microscope (AFM).
In what follows, we use the measurement data for
parallel corrugation axes. The plate was the diffraction grating
with uniaxial sinusoidal corrugations of period
$\Lambda=570.5\,$nm and amplitude $A_1=40.2\,$nm.
The side of the plate was equal to 1\,cm allowing to consider it
infinitely large. The grating was made of a hard epoxy
with density $\rho_e=1.08\,\mbox{g/cm}^3$ and coated with an Au
layer of thickness $\Delta_{\rm Au}^{\!(1)}=300\,$nm.
The corrugations on the bottom surface of a polystyrene sphere
of density $\rho_p=1.06\,\mbox{g/cm}^3$ were imprinted under
pressure from the plate.
For technological purposes the sphere was coated with a layer of
Cr of thickness $\Delta_{\rm Cr}^{\!(2)}=10\,$nm, then with a
layer of Al of thickness $\Delta_{\rm Al}^{\!(2)}=20\,$nm,
and finally with a layer of Au of thickness
$\Delta_{\rm Au}^{\!(2)}=110\,$nm.
The outer radius of the coated sphere was measured to be
$R=99.6\,\mu$m. The imprinted corrugations on the sphere have had
the same period as on the plate and the amplitude $A_2=14.6\,$nm.
The size of the imprint area was
$L_x\approx L_y\approx 14\,\mu$m,
i.e., much larger than the corrugation period $\Lambda$.

\subsection{Calculation of the normal force due to
exchange of two axion-like particles}

Now, let us calculate the normal (i.e., perpendicular to the mean
levels of boundary surfaces) force acting between a sinusoidally
corrugated spherical envelope of thickness $d_2$
 and a sinusoidally corrugated plate of thickness $d_1$ due to
exchange of two axion-like particles
between protons and neutrons belonging to
them. We assume the pseudoscalar character of an axion-fermion
coupling and neglect the axion-electron interaction \cite{26}.
The account of the latter and possible scalar axion-fermion
coupling \cite{50} could lead to only a minor strengthening of
the obtained constraints.

The pseudoscalar interaction of axions $a$ with fermions $\psi$
is described by the Lagrangian
\begin{equation}
{\cal L}_{\rm ps}=-i g_{ak}\bar{\psi}\gamma_5\psi a,
\label{0a}
\end{equation}
\noindent
where $g_{ak}$ is the coupling constant for an axion to a
proton ($k=p$) or to a neutron ($k=n$).
Note that the originally introduced QCD axions are pseudo
Nambu-Goldstone bosons and their interaction with fermions is
described by the pseudovector interaction Lagrangian
\begin{equation}
{\cal L}_{\rm pv}=\frac{g_{ak}}{2m_a}\bar{\psi}\gamma_5\gamma_{\mu}
\psi\partial^{\mu} a.
\label{0b}
\end{equation}
\noindent
Both the pseudoscalar and pseudovector Lagrangians (\ref{0a}) and
(\ref{0b})  result in one and the
same action after the integration by parts is performed.
However, the quantum field theory with the pseudoscalar Lagrangian
(\ref{0a}) is renormalizable, whereas it is nonrenormalizable
with the pseudovector Lagrangian (\ref{0b}). This may result in
some nonequivalence \cite{7,9}. Below we assume the pseudoscalar
interaction (\ref{0a}) which is applicable to a wide class of
axion models, specifically, to all GUT axion-like particles
\cite{50}.

First, we consider two plane parallel plates of infinitely large
area with thicknesses $d_1$ and $d_2$. Let the coordinate plane
$x,\,y$ coincide with an upper surface of the lower plate, and the
$z$ axis be perpendicular to it. Let $z$ be the varying
separation distance between the plates. Using the pseudoscalar
Lagrangian (\ref{0a}), the effective interaction potential due to
exchange of two axion-like particles
between two nucleons situated at the points
$\mbox{\boldmath$r$}_1$ of the lower plate and
$\mbox{\boldmath$r$}_2$ of the upper plate has the form
\cite{26,51,52}
\begin{equation}
V_{kl}(|\mbox{\boldmath$r$}_1-\mbox{\boldmath$r$}_2|)=-
\frac{g_{ak}^2g_{al}^2}{32\pi^3m^2}\,
\frac{m_a}{(\mbox{\boldmath$r$}_1-\mbox{\boldmath$r$}_2)^2}\,
K_1(2m_a|\mbox{\boldmath$r$}_1-\mbox{\boldmath$r$}_2|).
\label{eq1}
\end{equation}
\noindent
Here, $k,\,l=p$ for protons and $k,\,l=n$ for
neutrons, $m=(m_p+m_n)/2$ is the mean nucleon mass,
and $K_1(z)$ is the modified Bessel function of the second kind.
Equation (\ref{eq1}) is valid under the assumption
$|\mbox{\boldmath$r$}_1-\mbox{\boldmath$r$}_2|\gg1/m$
which is fulfilled with a large safety margin due to the fact
that in measurements of Ref.~\cite{47} the interacting bodies
are separated by more than 127\,nm.

The interaction energy per unit area of two plane parallel
plates due to two-axion exchange is given by
\begin{eqnarray}
&&
E(z)=2\pi\sum_{k,l}n_{k,1}n_{l,2}\int_{z}^{z+d_1}\!\!\!dz_1
\int_{-d_2}^{0}\!\!\!dz_2\int_{0}^{\infty}\!\!\!\rho d\rho
\nonumber \\
&&~~~~~~~~~~~~~~
\times
V_{kl}(\sqrt{\rho^2+(z_1-z_2)^2}),
\label{eq2}
\end{eqnarray}
\noindent
where $V_{kl}$ is defined in Eq.~(\ref{eq1}) and
\begin{equation}
n_{p,i}=\frac{\rho_i}{m_{\rm H}}\,
\frac{Z_i}{\mu_i}, \qquad
n_{n,i}=\frac{\rho_i}{m_{\rm H}}\,
\frac{N_i}{\mu_i}.
\label{eq3}
\end{equation}
\noindent
The index $i=1,\,2$ here numerates the lower and upper plates,
respectively, $\rho_{1,2}$ are the densities of plate materials,
$Z_{1,2}$ and $N_{1,2}$ are the numbers of protons and the mean
numbers of neutrons in the atoms (molecules) of the
plates. The quantities $\mu_{1,2}$ are defined as
$\mu_{1,2}=m_{1,2}/m_{\rm H}$, where $M_{1,2}$ and $m_{\rm H}$
are the mean masses of the atoms (molecules) of the plate
materials and the mass of the atomic hydrogen, respectively.
In Ref.~\cite{53} one can find the tabulated values of $Z/\mu$
and $N/\mu$ for the first 92 elements of the Periodic Table
taking into account their isotopic composition.

Substituting Eqs.~(\ref{eq1}) and (\ref{eq3}) into
Eq.~(\ref{eq2}), one obtains
\begin{eqnarray}
&&
E(z)=-\frac{m_a}{m^2m_{\rm H}^2}C_1C_2\int_{z}^{z+d_1}\!\!\!dz_1
\int_{-d_2}^{0}\!\!\!dz_2\int_{0}^{\infty}\!\!\!\rho d\rho
\nonumber \\
&&~~~~~~~~~~~~~~
\times
\frac{K_1(2m_a\sqrt{\rho^2+(z_1-z_2)^2})}{\rho^2+(z_1-z_2)^2},
\label{eq4}
\end{eqnarray}
\noindent
where
\begin{equation}
C_i=\rho_i\left(\frac{g_{ap}^2}{4\pi}\,\frac{Z_i}{\mu_i}+
\frac{g_{an}^2}{4\pi}\,\frac{N_i}{\mu_i}\right).
\label{eq5}
\end{equation}
\noindent
It is convenient to use the integral representation \cite{54}
\begin{equation}
\frac{K_1(t)}{t}=\int_{1}^{\infty}\!\!\!du\sqrt{u^2-1}e^{-tu}
\label{eq6}
\end{equation}
\noindent
and the new variable $v=\sqrt{\rho^2+(z_1-z_2)^2}$.
Then Eq.~(\ref{eq4}) can be rearranged to
\begin{eqnarray}
&&
E(z)=-\frac{C_1C_2}{4m_am^2m_{\rm H}^2}
\int_{1}^{\infty}\!\!\!du\frac{\sqrt{u^2-1}}{u^3}e^{-2m_auz}
\nonumber \\
&&~~~~
\times(1-e^{-2m_aud_1})(1-e^{-2m_aud_2}).
\label{eq7}
\end{eqnarray}

Now we assume that the opposite sides of
the plates under consideration are
covered with uniaxial sinusoidal corrugations of
equal period
$\Lambda$. There was no phase shift between corrugations on
both bodies in the experiment of Ref.~\cite{47}.
The corrugation axes are directed along the axis $y$.
Then, the separation distance between the surfaces of the
plates is given by
\begin{equation}
z\equiv z(x)=a+(A_1-A_2)\cos\frac{2\pi x}{\Lambda},
\label{eq8}
\end{equation}
\noindent
where $a$ is the separation distance between the mean levels
of corrugations on both plates. The interaction energy
due to two-axion exchange per unit area of corrugated plates
can be obtained by the method of geometrical averaging
\cite{55,56}
\begin{equation}
E_{\rm corr}(a)=\frac{1}{\Lambda}
\int_{0}^{\Lambda}\!\!\!dx E\left(z(x)\right),
\label{eq9}
\end{equation}
\noindent
where $E(z)$ and $z(x)$ are defined in Eqs.~(\ref{eq7}) and
(\ref{eq8}), respectively. For the Casimir forces,
Eq.~(\ref{eq9}) is an approximate one and works good under the
condition $\Lambda\gg a$ \cite{48,49}.
However, for the Yukawa-type forces and forces due to
two-axion exchange it is, in fact, exact.
Substituting Eqs.~(\ref{eq7}) and (\ref{eq8}) in Eq.~(\ref{eq9}),
one arrives at
\begin{eqnarray}
&&
E_{\rm corr}(a)=-\frac{C_1C_2}{4m_am^2m_{\rm H}^2\Lambda}
\int_{0}^{\Lambda}\!\!\!dx
\int_{1}^{\infty}\!\!\!du\frac{\sqrt{u^2-1}}{u^3}
\nonumber \\[1mm]
&&~~~~~
\times e^{-2m_aua} e^{-2m_au(A_1-A_2)\cos(2\pi x/\Lambda)}
\nonumber \\[1mm]
&&~~~~~
\times (1-e^{-2m_aud_1})(1-e^{-2m_aud_2}).
\label{eq10}
\end{eqnarray}
\noindent
The integration with respect to $x$ results in
\begin{eqnarray}
&&
E_{\rm corr}(a)=-\frac{C_1C_2}{4m_am^2m_{\rm H}^2}
\int_{1}^{\infty}\!\!\!du\frac{\sqrt{u^2-1}}{u^3}
\nonumber \\[1mm]
&&~~~~~
\times e^{-2m_aua} I_0\left(2m_au(A_1-A_2)\right)
\nonumber \\[1mm]
&&~~~~~
\times
(1-e^{-2m_aud_1})(1-e^{-2m_aud_2}),
\label{eq11}
\end{eqnarray}
\noindent
where $I_0(z)$ is the Bessel function of imaginary argument.

Finally, we should take into account that in the experiment of
Ref.~\cite{47} the upper test body was not a sinusoidally
corrugated plate, but a sinusoidally corrugated sphere.
The force $F_{\rm corr}^{(s,p)}$ acting between a corrugated
spherical envelope of thickness $d_2$ and a corrugated plate
can be expressed via the energy per unit area of two
corrugated plates (\ref{eq11}) by means of the proximity
force approximation (PFA) \cite{57,58}
\begin{equation}
F_{\rm corr}^{(s,p)}(a)=2\pi RE_{\rm corr}(a).
\label{eq12}
\end{equation}
\noindent
The PFA is sufficiently exact under the condition $a\ll R$
(in this case its error is less than $a/R$ for the Casimir
force \cite{59,60,61,62} and takes much less values for the
exponentially decreasing forces, such as the Yukawa force and
the force due to two-axion exchange).
One more condition for the validity of Eq.~(\ref{eq12}) in
application to forces due to the exchange of scalar or
pseudoscalar particles in that the Compton wavelength of a
hypothetical particle should be much less than the sphere
radius, i.e., $1/m_a\ll R$ \cite{63,64}. This is also
satisfied in our case taking into account that the
competitive constraints from the experiment of Ref.~\cite{47}
follow for $m_a>1\,$eV ($1/m_a<200\,$nm).
For the validity of Eq.~(\ref{eq12}) in the case of a spherical
envelope, its thickness should be sufficiently small, i.e.,
$d_2\ll R$ \cite{63,64}. As a result, Eqs.~(\ref{eq11}) and
(\ref{eq12}) provide an analytic expression for the force due
to the exchange of two axion-like particles
acting between a sinusoidally corrugated
spherical envelope of radius $R$ and thickness $d_2$ and
a sinusoidally corrugated plate of thickness $d_1$.

\subsection{Derivation of constraints}

Now we take into account that in the experiment of Ref.~\cite{47}
the plate was coated with an Au layer and the polystyrene sphere
was coated with layers of Cr, Al and Au, with the densities
$\rho_{\rm Cr}=7.15\,\mbox{g/cm}^3$,
$\rho_{\rm Al}=2.7\,\mbox{g/cm}^3$, and
$\rho_{\rm Au}=19.29\,\mbox{g/cm}^3$, respectively.
Note that the hard epoxy and polystyrene are materials of low
density and lead to negligibly small contributions to the force
due to exchange of two axion-like particles. By applying Eqs.~(\ref{eq11}) and
(\ref{eq12})
to each pair of metallic layers (the inequalities
$\Delta_{\rm Cr}^{\!(2)},\,\Delta_{\rm Al}^{\!(2)},\,
\Delta_{\rm Au}^{\!(2)}\ll R$ are satisfied with a large safety
margin), the resulting expression for the force due to two-axion
exchange in the experimental configuration of Ref.~\cite{47}
takes the form
\begin{eqnarray}
&&
F_{\rm corr}^{(s,p)}(a)=-\frac{\pi RC_{\rm Au}}{2m_am^2m_{\rm H}^2}
\int_{1}^{\infty}\!\!\!du\frac{\sqrt{u^2-1}}{u^3}e^{-2m_aua}
\nonumber \\[1mm]
&&~~~~~
\times  I_0\left(2m_au(A_1-A_2)\right)(1-e^{-2m_au\Delta_{\rm Au}^{\!(1)}})
\nonumber \\[1mm]
&&~~~~~
\times\left[C_{\rm Au}+(C_{\rm Al}-C_{\rm Au})
e^{-2m_au\Delta_{\rm Au}^{\!(2)}}\right.
\nonumber \\[1mm]
&&~~~~~~~
+(C_{\rm Cr}-C_{\rm Al})
e^{-2m_au(\Delta_{\rm Au}^{\!(2)}+\Delta_{\rm Al}^{\!(2)})}
\nonumber \\[1mm]
&&~~~~~~~
\left.
-C_{\rm Cr}
e^{-2m_au(\Delta_{\rm Au}^{\!(2)}+\Delta_{\rm Al}^{\!(2)}
+\Delta_{\rm Cr}^{\!(2)})}\right].
\label{eq13}
\end{eqnarray}
\noindent
Here, the coefficients $C_{\rm Au}$, $C_{\rm Al}$, and
$C_{\rm Cr}$ are defined according to Eq.~(\ref{eq5}),
as applied to respective elements. The quantities
$Z/\mu$ and $N/\mu$ are equal to \cite{53}
0.40422 and 0.60378,
0.48558 and 0.52304,
0.46518 and 0.54379
for Au, Al, and Cr, respectively.

The Casimir force between a corrugated sphere and a
corrugated plate was measured at separations $a\geq 127\,$nm
and found to be in good agreement with theoretical predictions
of the scattering theory \cite{47}. This theory is a generalization
of the Lifshitz theory of the van der Waals and Casimir forces
for the case of boundary surfaces of arbitrary shape \cite{65}.
It should be noted that there is a problem in
theory-experiment comparison connected with different models
for the low-frequency behavior of the dielectric permittivity
\cite{57,58}.
It turned out that the measurement data of the most precise
experiments using a metal-coated sphere above a metal-coated
plate agree with theoretical predictions only if the relaxation
of conduction electrons is omitted (in so doing the relaxation
properties of bound electrons are taken into account).
At separation distances of a few hundred nanometers the effect
of relaxation of free electrons comes to only several percent.
However, the theoretical predictions taking this kind of
ralaxation into account were reliably excluded in the dynamic
experiments performed by means of AFM \cite{37,38,39,40,41}
and micromachined oscillator \cite{42,43}.

Luckily, this problem is unrelated to experiments with corrugated
surfaces under consideration here.
The point is that they are performed by means of AFM at shorter
separation distances where the dymanic mode is inoperable.
As a result, in the static mode, the experimental error in force
measurements exceeds the theoretical uncertainty connected with
an account or neglect of the relaxation properties of free
charge carriers.

As was noted above, the measurement data of Ref.~\cite{47} were
found to be in a good agreement with theoretical predictions for
the Casimir force.
No signature of any other interaction, specifically,
due to exchange of two axion-like particles
was found in the limits
of the total experimental error, $\Delta F_C^{(s,p)}(a)$,
in the measured Casimir force. This error was determined at
the 67\% confidence level as a combination of random and
systematic errors. Thus, one can conclude
that any additional force between
the sphere and the plate, specifically, arising due to two-axion exchange
should satisfy the inequality
\begin{equation}
|F_{\rm corr}^{(s,p)}(a)|\leq\Delta F_C^{(s,p)}(a),
\label{eq14}
\end{equation}
\noindent
where the left-hand side is given in Eq.~(\ref{eq13}).
We have analyzed Eq.~(\ref{eq14}) numerically and found
that the strongest constraints on the coupling constants
$g_{an}$, $g_{ap}$ as functions of $m_a$ follow at the
shortest separation $a=127\,$nm. At this separation
distance the total experimental error determined at the
67\% confidence level is
$\Delta F_C^{(s,p)}(a)=0.94\,$pN \cite{47}.
Then, the constraints following from Eq.~(\ref{eq14}) are also
determined at the same 67\% confidence level.
Note that the constraints obtained in this way are the conservative
ones. The inclusion of any unaccounted attractive force between
a sphere and a plate, in addition to the force due to two-axion
exchange, could make them only stronger.

In Fig.~1(a) we show the constraints on the constants
$g_{ap(n)}^2/(4\pi)$ as functions of $m_a$ obtained from
Eq.~(\ref{eq14}) at $a=127\,$nm.
The three lines from bottom to top
corresponding to the equality sign in Eq.~(\ref{eq14})
are plotted under the
conditions $g_{ap}^2=g_{an}^2$, $g_{an}^2\gg g_{ap}^2$,
and $g_{ap}^2\gg g_{an}^2$, respectively.
The regions  of the plane above each line
are prohibited by the measurement results and
below each line are allowed.
The range of  masses of axion-like particles
from 1\,eV to 20\,eV is chosen.
For larger $m_a$ the strength of the constraints shown in
Fig.~1(a) decreases quickly. This is in line with the
constraints on the Yukawa-type interactions obtained in
Ref.~\cite{56} from the same experiment.
The strongest constraints on the Yukawa interaction
constant were found in the interaction range from 15 to
65\,nm \cite{56}, which corresponds to the mass of a hypothetical
scalar particle from 2 to 15\,eV.

In Fig.~1(b) the constraints derived here under the condition
$g_{ap}^2=g_{an}^2$ [the black line reproducing the lower line
of Fig.~1(a)] are compared with the strongest constraints
obtained \cite{35} under the same condition from measuring the
Casimir pressure by means of micromachined oscillator
\cite{42,43} (the gray line). As is seen in Fig.~1(b), the
constraints obtained here are stronger in the region of
masses of axion-like particles $m_a\geq 11\,$eV.

\section{Constraints from measuring the lateral Casimir force
between sinusoidally corrugated surfaces}

In Refs.~\cite{48,49} the lateral Casimir force was measured
by means of AFM as
a function of the phase shift $\varphi_0$ between uniaxial
corrugations of period $\Lambda=574.4\,$nm on the sphere and on
the plate over the region of separations from $a=120\,$nm to
$a=190\,$nm. The corrugation amplitudes on the plate and on the
sphere were $A_1=85.4\,$nm and $A_2=13.7\,$nm, respectively.
The top of the plate (the grating made of a hard epoxy) was
covered with $\Delta_{\rm Au}^{\!(1)}=300\,$nm Au coating.
The sphere was made of polystyrene and coated with a layer
of Cr of $\Delta_{\rm Cr}^{\!(2)}=10\,$nm and then with a
layer of Au of $\Delta_{\rm Au}^{\!(2)}=50\,$nm thickness.
The outer radius of the coated sphere was measured to be
$R=97.0\,\mu$m. Unlike the experiment of Ref.~\cite{47}
considered in Sec.~II, the corrugation axes on the plate and
on the sphere were always parallel, but the phase shift between
corrugations was varied leading to a nonzero lateral Casimir
force.

\subsection{Calculation of the lateral force due to
exchange of two axion-like particles}

Now we calculate the lateral force (acting in tangential
directions to the mean levels of boundary surfaces)
between the sinusoidally corrugated spherical envelope of
thickness $d_2$ and the sinusoidally corrugated plate of
thickness $d_1$, which arises due to two-axion exchange.

Similarly to Sec.~IIA, we consider first two plane parallel
plates with thicknesses $d_1$ and $d_2$ and again arrive at
the interaction energy per unit area due to two-axion
exchange given by Eq.~(\ref{eq7}). Then we assume that the
opposite sides of the plates are covered with uniaxial
sinusoidal corrugations of an equal period but with some
phase shift $\varphi=2\pi x_0/\Lambda$.
As a result, the separation distance between the surfaces
of the plates is given not by Eq.~(\ref{eq8}), but by the
following expressions:
\begin{equation}
z\equiv z(x)=a+A_1\cos\frac{2\pi x}{\Lambda}-
A_2\cos\frac{2\pi(x+x_0)}{\Lambda}.
\label{eq15}
\end{equation}
\noindent
Equation (\ref{eq15}) can be identically rearranged to
\begin{equation}
z=a+b\cos\left(\frac{2\pi x}{\Lambda}+\tilde{\varphi}
\right),
\label{eq16}
\end{equation}
\noindent
where
\begin{eqnarray}
&&
b=(A_1^2+A_2^2-2A_1A_2\cos\varphi)^{1/2},
\nonumber \\[1mm]
&&{\cot}\tilde{\varphi}=
\frac{A_1-A_2\cos\varphi}{A_2\sin\varphi}.
\label{eq17}
\end{eqnarray}

Substituting Eqs.~(\ref{eq7}) and (\ref{eq16}) in Eq.~(\ref{eq9})
and integrating with respect to $x$, we obtain
\begin{eqnarray}
&&
E_{\rm corr}(a)=-\frac{C_1C_2}{4m_am^2m_{\rm H}^2}
\int_{1}^{\infty}\!\!\!du\frac{\sqrt{u^2-1}}{u^3}
\nonumber \\[1mm]
&&~~~~~
\times e^{-2m_aua} I_0\left(2m_aub\right)
\nonumber \\[1mm]
&&~~~~~
\times
(1-e^{-2m_aud_1})(1-e^{-2m_aud_2}).
\label{eq18}
\end{eqnarray}
\noindent
This is the generalization of Eq.~(\ref{eq11}) for the case of
a nonzero phase shift between corrugations on both surfaces.

Using the PFA in Eq.~(\ref{eq12}), we now find the normal
force due to exchange of two axion-like particles
acting between a sinusoidally
corrugated spherical envelope of thickness $d_2$ and a
sinusoidally corrugated plate of thickness $d_1$
\begin{eqnarray}
&&
F_{\rm corr}^{(s,p)}(a)=-\frac{\pi RC_1C_2}{2m_am^2m_{\rm H}^2}
\int_{1}^{\infty}\!\!\!du\frac{\sqrt{u^2-1}}{u^3}
\nonumber \\[1mm]
&&~~~~~
\times e^{-2m_aua} I_0\left(2m_aub\right)
\nonumber \\[1mm]
&&~~~~~
\times
(1-e^{-2m_aud_1})(1-e^{-2m_aud_2}).
\label{eq19}
\end{eqnarray}
\noindent
This approximate expression is applicable under the conditions
$a\ll R$, $1/m_a\ll R$, and $d_2\ll R$ (see Sec.~IIA).

By the negative integration of Eq.~(\ref{eq19}) with respect to
$a$, one can obtain the interaction energy due to two-axion
exchange between the corrugated sphere and the corrugated plate
\begin{eqnarray}
&&
E_{\rm corr}^{(s,p)}(a)=-\frac{\pi RC_1C_2}{4m_a^2m^2m_{\rm H}^2}
\int_{1}^{\infty}\!\!\!du\frac{\sqrt{u^2-1}}{u^4}
\nonumber \\[1mm]
&&~~~~~
\times e^{-2m_aua} I_0\left(2m_aub\right)
\nonumber \\[1mm]
&&~~~~~
\times
(1-e^{-2m_aud_1})(1-e^{-2m_aud_2}).
\label{eq20}
\end{eqnarray}
\noindent
Then, the lateral force due to exchange of two axion-like particles
is obtained by
the negative differentiation of Eq.~(\ref{eq20}) with respect to
the phase shift between corrugations
\begin{eqnarray}
&&
F_{\rm corr,\,lat}^{(s,p)}(a)=
-\frac{\partial E_{\rm corr}^{(s,p)}(a)}{\partial x_0}=
-\frac{2\pi}{\Lambda}\,
\frac{\partial E_{\rm corr}^{(s,p)}(a)}{\partial\varphi}
\nonumber \\[1mm]
&&~~~
=\frac{\pi^2 RC_1C_2}{m_am^2m_{\rm H}^2}\,\frac{A_1A_2}{b\Lambda}
\sin\varphi
\int_{1}^{\infty}\!\!\!du\frac{\sqrt{u^2-1}}{u^3}
\nonumber \\[1mm]
&&~~~~~
\times e^{-2m_aua} I_1\left(2m_aub\right)
\nonumber \\[1mm]
&&~~~~~
\times
(1-e^{-2m_aud_1})(1-e^{-2m_aud_2}).
\label{eq21}
\end{eqnarray}
\noindent
This expression can be used for constraining the parameters of
axion-like particles from the experimental data of
Refs.~\cite{48,49}.

\subsection{Constraints from measuring the lateral Casimir
force}

To obtain the constraints on axion-like particles
from the experimental data,
it is necessary to take into account the layer structure of the
test bodies. Similarly to Sec.~IIB, the hard epoxy and polystyrene
lead to negligibly small contributions to the force due to
two-axion exchange. Thus, one should take into account the Au
layer on the plate, and Cr and Au layers on the
sphere. The resulting lateral force is found by applying
Eq.~(\ref{eq21}) two times.
We also take into account that the largest magnitude of the
lateral force due to two-axion exchange and, as a consequence,
strongest constraints are obtained for the phase shift
$\varphi=\pi/2$, $b=\sqrt{A_1^2+A_2^2}$. In this case
\begin{eqnarray}
&&
\max|F_{\rm corr,\,lat}^{(s,p)}(a)|=
\frac{\pi^2 RC_{\rm Au}}{m_am^2m_{\rm H}^2}\,
\frac{A_1A_2}{\Lambda\sqrt{A_1^2+A_2^2}}
\nonumber \\[1mm]
&&~~
\times
\int_{1}^{\infty}\!\!\!du\frac{\sqrt{u^2-1}}{u^3}
e^{-2m_aua} I_1\left(2m_au\sqrt{A_1^2+A_2^2}\right)
\nonumber \\[1mm]
&&~~~~~
\times
(1-e^{-2m_au\Delta_{\rm Au}^{\!(1)}})
\left[
\vphantom{e^{-2m_au\Delta_{\rm Au}^{\!(1)}}}
C_{\rm Au}+(C_{\rm Cr}-C_{\rm Au})
\right.
\nonumber \\[1mm]
&&~~~~\left.
\times e^{-2m_au\Delta_{\rm Au}^{\!(2)}}
-C_{\rm Cr}
e^{-2m_au(\Delta_{\rm Au}^{\!(2)}+\Delta_{\rm Cr}^{\!(2)})}
\right].
\label{eq22}
\end{eqnarray}
\noindent

Similarly to the experiment of Ref.~\cite{47} discussed in Sec.~II,
the measurement data for the lateral Casimir force were found to
be in a good agreement with theoretical predictions of the
scattering theory.
No additional lateral force was observed in Refs.~\cite{48,49} in
the limits of the total experimental error
 $\Delta F_{C,\,\rm lat}^{(s,p)}(a)$ in the measured lateral Casimir
force.
The theoretical uncertainty due to an account or neglect of the
relaxation properties of free charge carriers is again smaller
than this error.
Because of this, the
force due to exchange of two axion-like particles
is constrained by the inequality
\begin{equation}
\max|F_{\rm corr,\,lat}^{(s,p)}(a)|\leq
\Delta F_{C,\,\rm lat}^{(s,p)}(a),
\label{eq23}
\end{equation}
\noindent
where the left-hand side is given by Eq.~(\ref{eq22}).
The strongest constraints for axion-like particles with
$m_a\leq 20\,$eV are obtained from the measurement data at
$a=124.7\,$nm where the total experimental error is
$\Delta F_{C,\,\rm lat}^{(s,p)}(a)=4.7\,$pN \cite{48,49}.
This error was determined at a higher, 95\%, confidence
level. Thus, the reliability of the obtained constraints
is also characterized by the 95\% confidence level.

The constraints on the constants $g_{ap(n)}^2/(4\pi)$
following from Eq.~(\ref{eq23}) are shown in Fig.~2(a)
as functions of the axion mass.
The three lines from bottom to top
are plotted under the
conditions $g_{ap}^2=g_{an}^2$, $g_{an}^2\gg g_{ap}^2$,
and $g_{ap}^2\gg g_{an}^2$, respectively.
As in Fig.~1(a),
the regions  of the plane above each line
are prohibited  and
below each line are allowed.
The same range of masses of axion-like particles
$1\,\mbox{eV}\leq m_a\leq 20\,$eV is chosen.
For larger masses the strength of the obtained constraints
 quickly decreases.

In Fig.~2(b) we compare the constraints derived here under the condition
$g_{ap}^2=g_{an}^2$ [the black solid line reproducing the lower line
from Fig.~2(a)]  with the strongest constraints
derived in Sec.~IIB from measurements of the normal Casimir force
[the dashed line reproducing the lower line
from Fig.~1(a)] and derived
in Ref.~\cite{35}  from measurements of the
Casimir pressure by means of micromachined oscillator
[the gray solid line reproducing the similar line from Fig.~1(b)].
 As is seen in Fig.~2(b), the constraints obtained here
from measurements of the lateral Casimir force
are stronger than those from
measurements of the normal Casimir force for $m_a\geq 5\,$eV
and stronger than the constraints following from measurements of
the Casimir pressure by means of micromachined oscillator for
$m_a\geq 8\,$eV.

\section{Comparison between different model-independent
constraints}

Many constraints on the parameters of axions and axion-like
particles were obtained with the help of some model approaches.
As discussed in Sec.~I, within the model of hadronic axions,
whose coupling constant is a function of $m_a$, the upper limits
for the axion mass $m_a\leq 159\,$eV \cite{20} and
$m_a\leq 145\,$eV \cite{21} were obtained using the detector of
$\gamma$-quanta appearing in the deexcitation of the nuclear
level excited by solar axions.
The parameters of hadronic axions were also found from the
neutrino data of supernova SN~1987A \cite{22} and from
astrophysical arguments connected with stellar cooling by the
axion emission \cite{23,24}.
In so doing, some model description of dense nuclear matter was
used \cite{24}. Here, we collect the model-independent
constraints on the coupling constants of axion-like particles
to nucleons which were
obtained from table-top laboratory experiments and are relevant
to different models of axions.

In Fig.~3 the constraints on the coupling constant of
axion-like particles to neutrons
as functions of $m_a$ are shown over the wide range of axion
masses from $10^{-10}\,$eV to 20\,eV.
The solid line 1 in Fig.~3 shows the constraints obtained
\cite{25}
by means of a magnetometer using spin-polarized K and ${}^3$He
atoms. These constraints are found in the region
from $10^{-10}\,$eV to $6\times 10^{-6}\,$eV.
The constraints obtained \cite{31} from the modern Cavendish-type
experiment \cite{32} in the region
from $m_a=10^{-6}\,$eV to $6\times 10^{-2}\,$eV are shown by the solid
line 2. The dashed lines 3 and 4 indicate the weaker constraints
found in Ref.~\cite{26} from the older Cavendish-type experiments
\cite{29,30} and from the E\"{o}tvos-type experiment \cite{27},
respectively. These constraints extend
from $m_a=10^{-8}\,$eV to $m_a= 4\times 10^{-5}$ (line 3) and
to $10^{-5}\,$eV (line 4).
They are obtained under the assumption that $g_{an}=g_{ap}$.
All the constraints indicated by the lines 5--8 are also obtained
under this assumption.
The solid line 5 found in the region from $10^{-3}\,$ to 15\,eV
follows \cite{35} from measurements of the Casimir
pressure by means of micromachined oscillator,
the dashed line 6 was obtained \cite{34}
in the region from $3\times 10^{-5}\,$ to 1\,eV
from measuring the gradient of the Casimir force by means of
dynamic AFM \cite{37,38}, and the dashed line 7 was found
\cite{33} in the region from $10^{-4}\,$ to 0.3\,eV
from the experiment on measuring the Casimir-Polder force
\cite{36}. Finally, the solid line 8 indicates the constraints
obtained here in the region of masses of axion-like
particles from 1 to 20\,eV
from measuring the lateral Casimir force between corrugated
surfaces (as shown in Sec.~IIB, the experiment of Ref.~\cite{47}
on measuring the normal Casimir force in the same configuration
leads to slightly weaker constraints).
 The region  above each line in Fig.~3 is
prohibited by the results of respective experiment whereas the
region below each line is allowed.

As can be seen in Fig.~3, the strength of laboratory
constraints on the axion to nucleon coupling constants decreases
with increasing axion mass. The same is true for the Yukawa-type
corrections to Newtonian gravitational law which arise due to
exchange of light scalar particles between atoms and molecules
of interacting bodies (see Ref.~\cite{57} for a review and
Refs.~\cite{55,56,66,67,68,69} for the most recent results).
In fact, for both additional interactions (due to the exchange of
pseudoscalar and scalar particles, respectively) the
gravitational experiments of E\"{o}tvos and Cavendish type lead
to strongest constraints within the moderately short interaction
regions (moderately large masses). For shorter interaction
regions (larger masses) stronger constraints are obtained from
measurements of the Casimir and Casimir-Polder forces
\cite{55,56,66,67,68,69}. This is illustrated in Fig.~3, where
the constraints on  couplings of axion-like particles
to nucleons obtained from
the Casimir effect become stronger than those obtained from the
gravitational experiments for $m_a\geq 2\times 10^{-3}\,$eV.

There are other purely laboratory experiments setting strong
constraints on the coupling constants of axion-like
particles with larger $m_a$ to nucleons. Recently
such constraints have been obtained \cite{70} from the comparison
of nuclear magnetic resonance measurements with calculations of
the scalar spin-spin interaction in deuterated molecular
hydrogen. The combination of the coupling constants
$g_{ap}(g_{ap}+g_{an})/(4\pi)$ was shown to be less than
$3.6\times 10^{-7}$ for $m_a$ in the range of $5\times 10^3\,$eV.

\section{Conclusions and discussion}

In this paper, we have obtained constraints on the coupling
constants of axion-like particles to nucleons from measuring
the normal and lateral Casimir force between sinusoidally
corrugated Au-coated surfaces of a sphere and a plate
\cite{47,48,49} by means of AFM.
These constraints are obtained from the measure of agreement
between the measured and calculated Casimir forces.
It is worth noting that calculation of the Casimir interaction
between arbitrarily shaped test bodies is a complicated
theoretical problem which was solved recently using the
quantum-field theoretical formalism of functional determinants
and scattering amplitudes \cite{65}.
Taking into account that the test bodies in the experiments of
Refs.~\cite{47,48,49} are unpolarized, there is no any additional
force due to an exchange of one axion between protons and
neutrons.
There is, however, the additional force between corrugated
surfaces arising due to  exchange of two axion-like particles,
and it is constrained
by the magnitude of the experimental error in measurements of
the Casimir force.

We have calculated both the normal and lateral forces due to
exchange of two axion-like particles between protons and
neutrons of sinusoidally
corrugated surfaces of a sphere and a plate.
The respective model-independent constraints on the coupling
constants of axion-like particles to protons and to neutrons were obtained
for the region of masses from 1\,eV to 20\,eV.
It was shown that the experiment of Ref.~\cite{47} on
measuring the normal Casimir force leads to the strongest
constraints on the coupling constants $g_{ap(n)}$ for
masses  of axion-like particles $m_a\geq 11\,$eV.
In this region of $m_a$, the
constraints obtained
here are stronger than those following from
measurements of the effective Casimir pressure by means of
micromachined oscillator \cite{35}.
We have obtained even stronger constraints on the coupling
constants of axion-like particles to nucleons from the experiment of
Refs.~\cite{48,49}, where the lateral Casimir force between
sinusoidally corrugated surfaces has been measured.
These constraints are stronger than those found from
measurements using a micromachined oscillator in the region
of masses  of axion-like particles $m_a\geq 8\,$eV.

The obtained constraints were compared with other
laboratory constraints on the coupling constants of
axions to nucleons found in different experiments.
We have considered constraints following from the magnetometer
measurements with polarized atoms \cite{25}, from the
experiments of E\"{o}tvos type \cite{26,27,28}, different
experiments of Cavendish type \cite{26,29,30,32}, and from
various measurements of the Casimir interaction
\cite{36,37,38,42,43,48,49}. These constraints cover the
wide range of masses  of axion-like particles
from $10^{-10}\,$eV to 20\,eV.

In future, it seems promising to perform measurements of the
Casimir interaction between polarized (magnetized) test
bodies. In this case, the process of a one-axion exchange
between nucleons would contribute to the additional force and
result in much stronger constraints on the coupling
constants of axion-like particles to nucleons.

%%%%%%%%%%%%%%%%%%%%%%%%%%%%%%%%%%%%%%%%%%%%%%%%%%%%%%%%%%
\section*{Acknowledgments}

The authors of this work acknowledge CNPq (Brazil) for
 partial financial support.
G.L.K.\ and V.M.M.\ are grateful to U.~Mohideen for
useful information about his experiments and to
M.\ I.\ Eides for helpful discussions.
They also acknowledge the Department
of Physics of the Federal University of
Para\'{\i}ba (Jo\~{a}o Pessoa, Brazil) for hospitality.

%%%%%%%%%%%%%%%%%%%%%%%%%%%%

%%%%%%%%%%%%%%%%%%%%%%%%%%%
%%%%%%%__FIGURE__1__%%%%%%%%%%%%%%%%%%%%
\begin{figure}[b]
\vspace*{-3cm}
\centerline{\hspace*{2cm}
\includegraphics{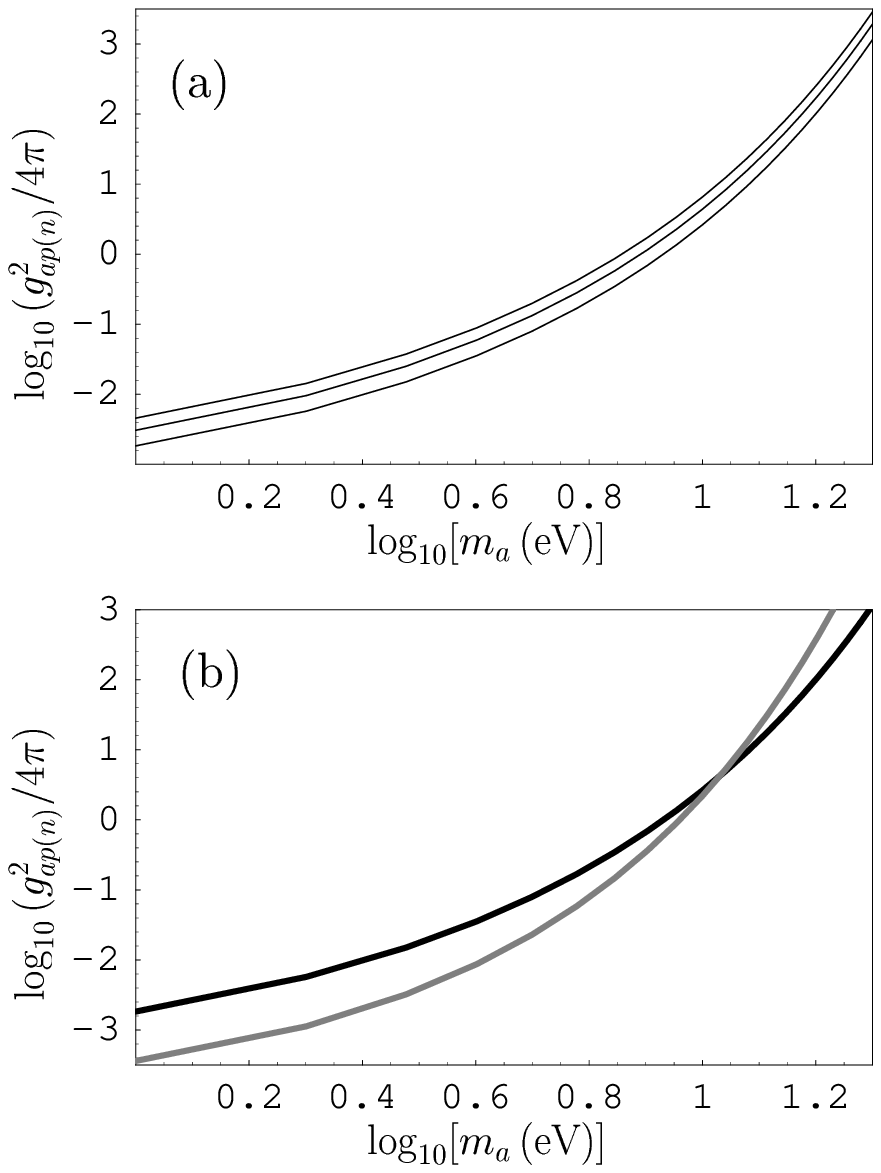}
}
\vspace*{-13cm}
\caption{
Constraints on the  coupling constants
of axion-like particles to nucleons are shown
as functions of the axion mass.
 The regions  above each line
are prohibited and below each line are allowed.
(a) Constraints from measurements of the normal
Casimir force between corrugated surfaces of a
sphere and a plate.
The lines from bottom to top are plotted under the
conditions $g_{ap}^2=g_{an}^2$, $g_{an}^2\gg g_{ap}^2$,
and $g_{ap}^2\gg g_{an}^2$, respectively.
(b) Constraints from measurements of the normal
Casimir force between corrugated surfaces of a
sphere and a plate and from the experiment using a
micromachined oscillator are plotted under the
condition $g_{ap}^2=g_{an}^2$ by the black and gray
lines, respectively.
}
\end{figure}
%%%%%%%%%%%%%
%%%%%%%__FIGURE__2__%%%%%%%%%%%%%%%%%%%%
\begin{figure}[b]
\vspace*{-3cm}
\centerline{\hspace*{2cm}
\includegraphics{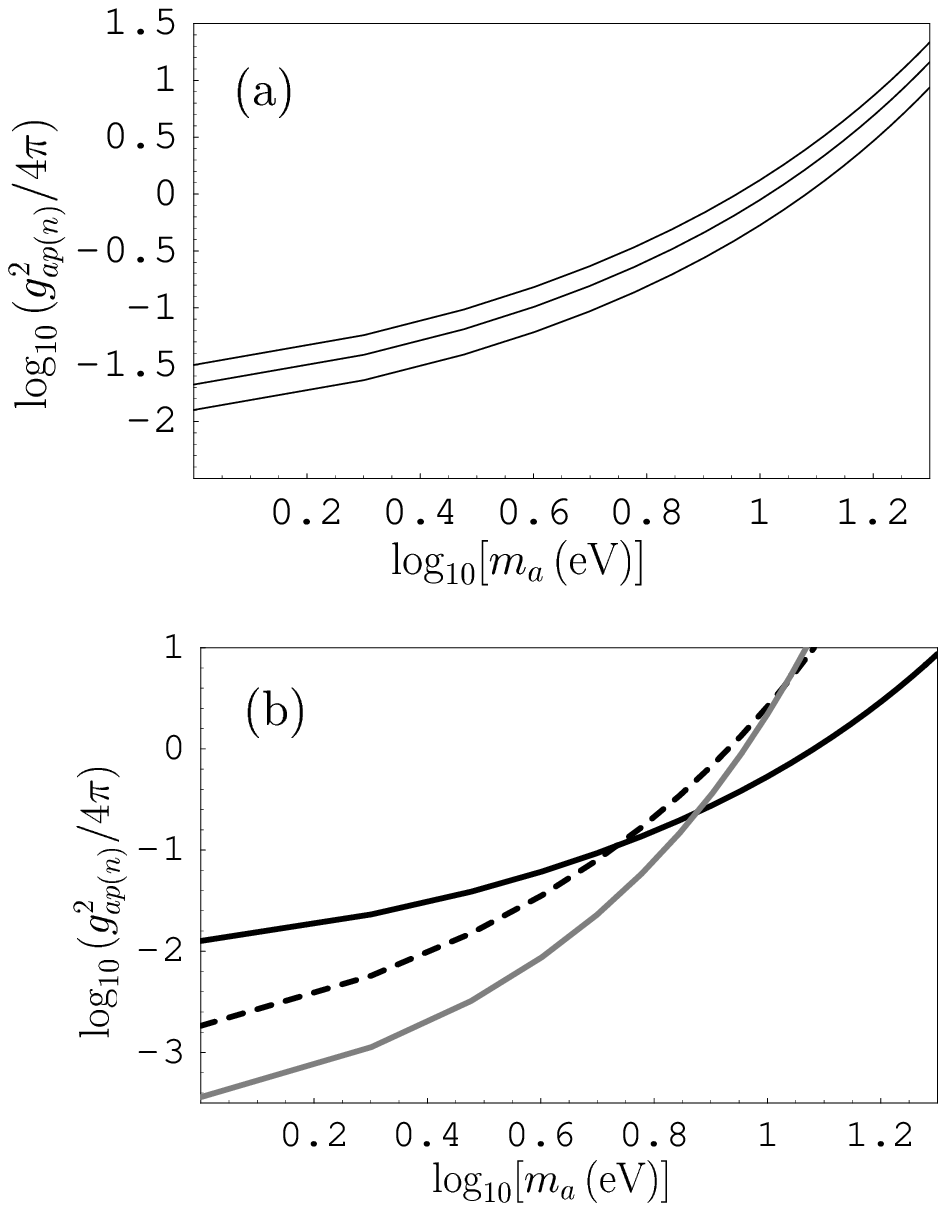}
}
\vspace*{-13cm}
\caption{
Constraints on the  coupling constants
of axion-like particles to nucleons are shown
as functions of the axion mass.
 The regions  above each line
are prohibited and below each line are allowed.
(a) Constraints from measurements of the lateral
Casimir force between corrugated surfaces of a
sphere and a plate.
The lines from bottom to top are plotted under the
conditions $g_{ap}^2=g_{an}^2$, $g_{an}^2\gg g_{ap}^2$,
and $g_{ap}^2\gg g_{an}^2$, respectively.
(b) Constraints from measurements of the lateral
and normal
Casimir force between corrugated surfaces of a
sphere and a plate are plotted under the
condition $g_{ap}^2=g_{an}^2$ by the solid and
dashed black lines, respectively.
The constraints from measuring the Casimir pressure
by means of micromachined oscillator are shown by
the gray line.
}
\end{figure}
%%%%%%%%%%%%%
%%%%%%%__FIGURE__3__%%%%%%%%%%%%%%%%%%%%
\begin{figure}[b]
\vspace*{-11cm}
\centerline{\hspace*{2cm}
\includegraphics{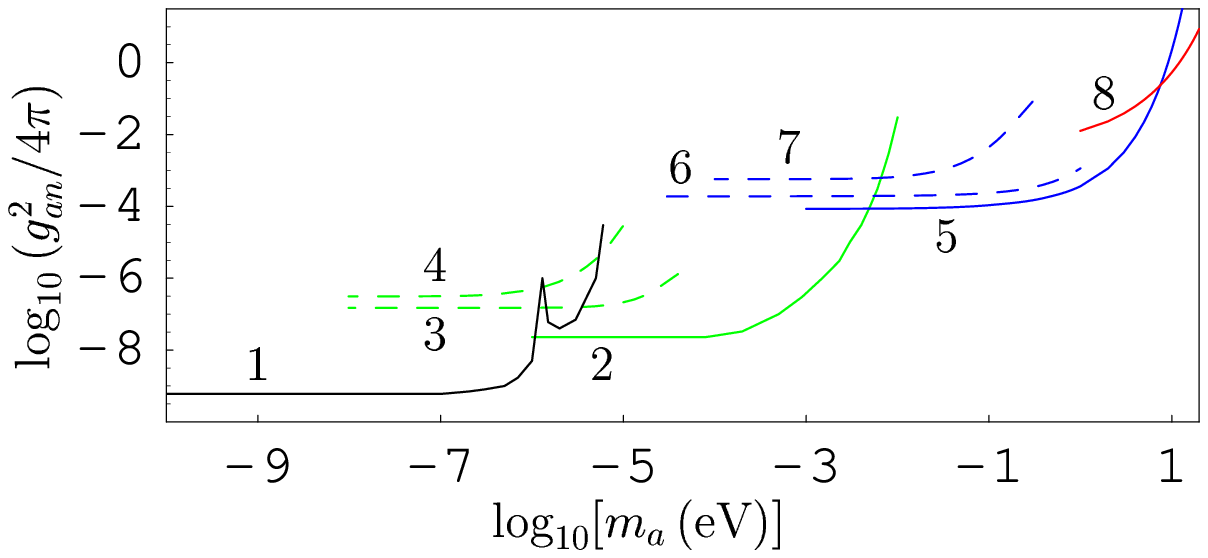}
}
\vspace*{-10cm}
\caption{(Color online) The laboratory
constraints on the  coupling constants
of axion-like particles to a neutron obtained from different
experiments are shown by the lines 1--8
as functions of the axion mass.
The line 1 follows from the magnetometer
measurements \cite{25}. The lines 2--4 are found
\cite{26,31} from the gravitational experiments
\cite{27,28,29,30,32}, and the lines 5--8 are obtained in
Refs.~\cite{33,34,35} and in this paper from the
Casimir effect \cite{36,37,38,42,43,48,49}
(see text for further discussion).
 The regions  above each line
are prohibited and below each line are allowed.
}
\end{figure}
%%%%%%%%%%%%%
\end{document}